\documentclass[onecolumn, floatfix, preprintnumbers, eqsecnum, letterpaper, superscriptaddress, nofootinbib]{revtex4}
\usepackage{graphicx}
\usepackage{microtype}
\usepackage{amsmath}
\usepackage{amssymb}
\usepackage{subfigure}
\usepackage{hyperref}
\usepackage{url}
\usepackage{xcolor}
\usepackage{color}
\usepackage{mathrsfs}
\usepackage{calrsfs}
\usepackage{amsfonts}
\usepackage{eufrak}
\usepackage{tabularx}
\usepackage{eucal}
\usepackage{latexsym}
\usepackage{ragged2e}
\usepackage{epsfig}
\usepackage{textcomp}
\usepackage{inputenc}

\usepackage{caption}
\usepackage{subfigure}

\usepackage{marvosym}
\usepackage{ifsym}

\usepackage{lipsum}

\DeclareCaptionJustification{justified}{\leftskip=0pt \rightskip=0pt \parfillskip=0pt plus 1fil}
\definecolor{vividviolet}{rgb}{0.62, 0.0, 1.0}
\definecolor{amaranth}{rgb}{0.9, 0.17, 0.31}
\definecolor{palatinateblue}{rgb}{0.15, 0.23, 0.89}
\definecolor{brightpink}{rgb}{1.0, 0.0, 0.5}
\definecolor{cornflowerblue}{rgb}{0.39, 0.58, 0.93}
\definecolor{deepcarminepink}{rgb}{0.94, 0.19, 0.22}
\definecolor{radicalred}{rgb}{1.0, 0.21, 0.37}

\hypersetup{ linktoc=all,
	colorlinks, linkcolor={palatinateblue},
	citecolor={brightpink}, urlcolor={amaranth}
}

\graphicspath{{Images/}}

\graphicspath{{Images/}}

\def\@fnsymbol#1{\ensuremath{\ifcase#1\or \ddagger \or  $\textleaf$  \or \dagger
		\else\@ctrerr\fi}}%

\makeatother

\def\sideremark#1{\ifvmode\leavevmode\fi\vadjust{\vbox to0pt{\vss
			\hbox to 0pt{\hskip\hsize\hskip1em
				\vbox{\hsize1.3cm\tiny\raggedright\pretolerance10000
					\noindent #1\hfill}\hss}\vbox to8pt{\vfil}\vss}}}%
\def\beq{\begin{equation}}
	\def\eeq{\end{equation}}

\begin{document}
\title{Gravitational lensing and shadow by a Schwarzschild-like black hole \\ in metric-affine bumblebee gravity}

\author{Xiao-Jun Gao}\email{gaoxiaojun@gznu.edu.cn}
\address{School of Physics and Electronic Science, Guizhou Normal University,
Guiyang 550001, People's Republic of China}
\address{College of Physics, Nanjing University of Aeronautics and Astronautics, Nanjing 211106, China}



\begin{abstract}
In this paper, we investigate the gravitational lensing effect and the shadow around a Schwarzschild-like black hole in metric-affine bumblebee gravity, which leads to the Lorentz symmetry breaking. We first present a generalized formalism for calculating higher-order corrections to light weak bending angle in a static, spherically symmetric and not asymptotically flat spacetime, and then applying this general formalism to the metric-affine bumblebee gravity. Moreover, we derive the light deflection angle and the size of the Einstein ring within the weak field in this scenario. In addition, we analyze the black hole shadow in this theory framework. By using observational data from the Einstein's ring of the galaxy ESO325-G004 and the black hole shadow of the ${\rm M}87$ galaxy, we estimate the upper bounds of the Lorentz symmetry breaking coefficient $\ell$, respectively.


\end{abstract}

\maketitle



\section{Introduction}
It is well-known that General Relativity (GR) is the standard theory of gravity, which has provided a remarkably successful explanation for gravitational phenomena at the classical level \cite{Will:2014kxa,LIGOScientific:2016aoc}. Without taking into account the quantum effects, GR preserves the Lorentz invariance. Nevertheless, when the quantization of the spacetime is considered in high-energy regime, the Lorentz symmetry breaking (LSB) could be generated \cite{cpt01}. Therefore, GR is thought to be inadequate for describing gravity and spacetime geometry effect at the quantum level.
It is generally thought that GR might need to be modified to form a unified gravity theory \cite{Clifton:2011jh}. In the pursuit of this unification some modified gravity theories have already been constructed, i.e. string theory \cite{Kostelecky:1988zi}, the loop quantum gravity \cite{Gambini:1998it,Bojowald:2004bb}, Einstein-æther theory \cite{Jacobson:2000xp}, Horava-Lifshitz gravity \cite{Horava:2009uw}.

Lorentz violation can be described by an effective field theory \cite{Kostelecky:1994rn}. Theories that violate Lorentz symmetry at the Planck scale, including both the GR and the SM, are encompassed by effective field theories called the standard-model extension (SME) \cite{Kostelecky:2003fs,Kostelecky:2020hbb}. It is worth mentioning that the SME constitutes a general effective field framework describing all
possible coefficients for Lorentz/CPT violation \cite{Kostelecky:2003fs}. Particularly,  the gravitational sector in the SME has been established on a Riemann-Cartan manifold, while the torsion was regarded as a dynamic geometrical quantity apart from the metric. Even though the non-Riemannian context has been considered for the gravity SME sector, most studies have mainly focused on the metric approach to gravity, in which the metric is treated as the merely dynamical geometric field. The vast majority of works investigate modified theories of gravity by using the usual metric approach, but it is still imperative to consider a more generic geometrical framework. In this context, the induction of gravitational topological terms \cite{Nascimento:2021vou} is one of the more relevant examples for investigating theories of gravity in a Riemann-Cartan background. Another  intriguing non-Riemannian geometry that has been investigated in the literature called the Finsler one \cite{BaoFinsler}, which has been extensively investigated in plenty of recent studies related to LSB \cite{Foster:2015yta,Edwards:2018lsn,Schreck:2014hga,Colladay:2015wra,Schreck:2015seb}.

The metric affine (Palatini) formalism has been thought of as the nontrivial generalization of the metric approach, where  the metric and connection are independent
dynamical geometrical quantities, and the corresponding discussion and some intriguing findings within the Palatini approach see \cite{Ghilencea:2020piz,Ghilencea:2020rxc}. Although there exist several recent works with respect to bumblebee gravity \cite{Delhom:2019wcm,Delhom:2020gfv,Delhom:2022xfo,Casana:2017jkc}, LSB in the Palatini approach framework has not constantly received much attention in the literature. Until recent instructive and interesting study \cite{Filho:2022yrk} fills up the gap by obtaining the exact solution for a particular metric-affine bumblebee gravity model. Spacetime properties of the above-mentioned solution have been extensively studied in \cite{Filho:2024isd,Filho:2024hri,Jha:2023vhn,Lambiase:2023zeo,Amarilo:2023wpn,Filho:2023etf}. In this present work, we will continually investigate the weak gravitational lensing (GL) and black hole shadow to deeply explore LSB leads to potential indicators of novel physical phenomena in the metric-affine bumblebee gravity model.

GL has been a powerful tool for investigating spacetime geometry \cite{Perlick:2004tq}. A few pioneering works have been done on the calculation of the light weak deflection angle by utilizing the parameterization-post-Newtonian (PPN) formalism. The first one was \cite{Keeton:2005jd}, where the light deflection angle was calculated in a static, spherically symmetric and asymptotically flat spacetime around a compact object, and later in \cite{Keeton:2006sa} the fundamental relations among the lensing observables (the image positions, magnifications, and time delays) were also calculated. This handle weak deflection angle method has been extensively applied \cite{Gao:2019pir,Gao:2023ltr,Zhang:2022nnj,Cheng:2021hoc,Gao:2021luq,Liu:2019pov,Hu:2013eya,QiQi:2023nex,Meng:2023wgi}. The light ray which passes close to a black hole would be deflected very strongly and even travel on circular orbits. This strong deflection \cite{Virbhadra:2022ybp,Virbhadra:2022iiy,Zhang:2023rsy,Gao:2021lmo} can lead to no light coming out of a black hole, and then the center of a black hole is seen as a dark region in the sky. The phenomenon is called black hole shadow. Based on numerical simulations, the seminal papers \cite{Falcke:1999pj,Melia:2001dy} predicted in 2000 that black hole shadow could actually be observed at wavelengths near 1mm with Very Long Baseline Interferometry (VLBI). Until 2019, the Event Horizon Telescope (EHT) Collaboration published that the shadow image of the black hole at the core of the ${\rm M}87$ galaxy \cite{EventHorizonTelescope:2019dse,EventHorizonTelescope:2019ths}. Inspired by this achievement, great attention is now focused on the investigation of various aspects of black hole shadows, see \cite{Haroon:2018ryd,Wei:2020ght,Perlick:2021aok,Guo:2022nto,Meng:2024puu,Meng:2023htc,Wang:2023vcv,Gao:2023mjb,Gralla:2019xty,Narayan:2019imo,Zeng:2022pvb,Zeng:2021mok,Zeng:2020vsj,Zeng:2020dco,Guo:2021bhr}.

For a static, spherically symmetric and asymptotically flat spacetime, the literature \cite{Keeton:2005jd} has derived a general formalism corrections to weak deflection angle in the PPN framework, where the spacetime metric can be expressed as a
series expansion in the single parameter $m_{\bullet}$ (the gravitational radius of the compact body), among zero-order term of the metric is assumed as one rather than an undetermined constant. In some modified gravity theories, if the nonvanishing of the Riemann tensor in
the limit the radial coordinate $r\rightarrow\infty$, the spacetime is not asymptotically
Minkowski. Therefore, the above assuming scenario is unsuitable for modified gravity model in a not asymptotically flat spacetime (e.g. the metric-affine bumblebee gravity model).
The zero-order term of the metric, which is expanded as a Taylor series in $m_{\bullet}$, need to be slightly modified definitions for a generic static, spherically symmetric and not asymptotically flat spacetime.  In additional, GL and shadow signatures of a Schwarzschild-like black hole in metric-affine bumblebee gravity \cite{Filho:2022yrk} are
barely known and relevant studies are still absent, requiring a detailed investigation.

As stated motivation, in this paper we first derive a generalized PPN formalism of the weak deflection angle in a static, spherically symmetric and not asymptotically flat spacetime background, and then we apply this general formalism to a Schwarzschild-like black hole in metric-affine bumblebee gravity. Meanwhile, we will further investigate the size of the Einstein ring and black hole shadow in this scenario, and compare our theoretical results
with observational data to estimate the Lorentz-violating coefficient $\ell$.

Our paper is organized as follows: In Sec.~\ref{section2}, we derive the generalized light weak deflection angle with the PPN formalism in a static, spherically symmetric and not asymptotically flat spacetime. In Sec.~\ref{section3}, applying the general formalism derived in Sec.~\ref{section2} to metric-affine bumblebee gravity, we obtain the actual form of light weak deflection angle and calculate the size of the Einstein's ring. We also provide the bounds to LSB
coefficient $\ell$ from astrophysics observational data of the Einstein's ring. In Sec.~\ref{section4}, we investigate the black hole shadow in this scenario and constrain the upper bound of the $\ell$ in terms of the observational data of the supermassive black hole shadow at the core of the ${\rm M}87$ galaxy. Finally Sec.~\ref{section5} is for summary and conclusion. Throughout this paper we use the geometric units with $G=c=1$ unless otherwise specified.

\section{Generalized formalism for light deflection angle}
\label{section2}
In a static, spherically symmetric and asymptotically Minkowski spacetime, Keeton and Petters have derived  a general PPN formalism for the light deflection angle in weak field limit \cite{Keeton:2005jd}, but it is unmerited to apply to some modified gravity models in the not asymptotically Minkowski spacetime. Therefore, in this section, the zero-order term of the metric with a small quantity expansion is redefined as an undetermined constant in a static, spherically symmetric and not asymptotically flat spacetime background, and we finally derive a generalized PPN formalism of the weak deflection angle.

The line element of a static four-dimensional spherically symmetric spacetime can be written as
\begin{align}
 ds^2=-A(r)dt^2+B(r)dr^2+r^2(d\theta^2+\sin^2\theta d\phi^2), \label{4Dspacetime}
\end{align}
in which we assume that $A(r)\rightarrow constant$ and $B(r)\rightarrow constant$ as $r\rightarrow \infty$ in a not asymptotically Minkowski spacetime. Without losing generality, we choose the equatorial plane $\theta=\pi/2$ to simply investigate the path of the light. One can write the conserved energy ($E$) and angular momentum ($L$) of light in the static spherically symmetric spacetime (\ref{4Dspacetime}) as follows \cite{Ishihara:2016vdc}:
\begin{align}
E=A(r)\dfrac{dt}{d\tau},~~L=r^2\dfrac{d\phi}{d\tau},\label{ELconstant}
\end{align}
where $\tau$ is the affine parameter along the light ray.
Utilizing the above equations, the light's trajectory equation along the equatorial plane is further obtained   \cite{Hu:2013eya,Ishihara:2016vdc}
\begin{align}
\dfrac{d\phi}{dr}=\dfrac{1}{r^2}\sqrt{\dfrac{A(r)B(r)}{1/b^2-A(r)/r^2}},\label{trajectory2}
\end{align}
where $b\equiv L/E$ is called the impact parameter. One can use $dr/d\phi|_{r=r_0}=0$ to obtain the relation between the minimum distance $r_0$ and the impact parameter, which satisfies
\begin{align}
b=\dfrac{r_0}{\sqrt{A(r_0)}}.\label{impactb}
\end{align}

If a light ray originates in the asymptotically region of spacetime, and it is deflected around compact body before arriving at an observer in the asymptotically region. In terms of trajectory (\ref{trajectory2}), the corresponding deflection angle of the light ray is expressed as
\begin{align}
\hat{\alpha}(r_0)=2\int_{r_0}^{\infty}\dfrac{1}{r^2}\sqrt{\dfrac{A(r)B(r)}{1/b^2-A(r)/r^2}}dr-\pi.\label{deflectionangle}
\end{align}
For convenience in later calculation, in terms of a new variable $x=r_0/r$, the equation (\ref{deflectionangle}) is rewritten as
\begin{align}
\hat{\alpha}(r_0)=2\int_{0}^{1}\sqrt{\dfrac{A(r_0/x)B(r_0/x)}{A(r_0)-A(r_0/x)x^2}}dx-\pi.\label{newdeflectionangle}
\end{align}
where we have used the equation (\ref{impactb}). From the (\ref{deflectionangle}) and (\ref{newdeflectionangle}), it is clearly seen the deflection angle $\hat{\alpha}(r_0)$ is determined by the metric components $A(r)$ and $B(r)$. The deflection angle in (\ref{deflectionangle}) or (\ref{newdeflectionangle}) as elliptic integral can not be
analytical evaluated particularly. The gravitational field outside of lens is assumed weak in order to handle this difficult. Therefore, one can expand the $A(r)$ and $B(r)$ as a series of the small parameter $M/r$, where $M$ is mass of a compact object. In the scenario, we can approximatively evaluate the integral of integrand term by term.

In the following, we calculate the generalized PPN formalism of the light weak deflection angle in a static, spherically symmetric and not asymptotically flat spacetime, where approximations up to the third order. To this end, we expand the $A(r)$ and $B(r)$ as a Taylor series to third order in $M/r\ll 1$ as follows
\begin{align}
A(r)=&a_0+a_1\dfrac{M}{r}+a_2\left(\dfrac{M}{r}\right)^2+a_3\left(\dfrac{M}{r}\right)^3+\mathcal{O}\left(\dfrac{M}{r}\right)^4,\label{Arexpand}\\
B(r)=&b_0+b_1\dfrac{M}{r}+b_2\left(\dfrac{M}{r}\right)^2+b_3\left(\dfrac{M}{r}\right)^3+\mathcal{O}\left(\dfrac{M}{r}\right)^4,\label{Brexpand}
\end{align}
where $a_i$ and $b_i$ denote PPN parameters. For the Schwarzschild metrics, they are written by
\begin{align}
a_0=1,~~a_1=-2,~~a_2=a_3=0;~~b_0=1,~~b_1=2,~~b_2=4,~~b_3=8,\label{SchwarPPN}
\end{align}
respectively. Here the $a_0=b_0=1$ for gravity theories in the asymptotically Minkowski
spacetime, but do not for not asymptotically flat spacetime. For example, we give the metric-affine bumblebee gravity to application in next section.

Next substituting (\ref{Arexpand}) into (\ref{impactb}), and then making a Taylor series to third order in $M/r_0\ll 1$ on the right hand
side of the (\ref{impactb})
\begin{align}
b=r_0\left[\dfrac{1}{\sqrt{a_0}}-\dfrac{a_1}{2\sqrt{a_0^3}}\dfrac{M}{r_0}+\dfrac{3a_1^2-4a_0a_2}{8\sqrt{a_0^5}}\left(\dfrac{M}{r_0}\right)^2-\dfrac{5a_1^3-12a_0a_1a_2+8a_0^2a_3}{16\sqrt{a_0^7}}\left(\dfrac{M}{r_0}\right)^3+\mathcal{O}\left(\dfrac{M}{r_0}\right)^4\right],\label{r0expb}
\end{align}
The radial coordinate distance $r_0$ needs to be replaced by $b$ to express the following deflection angle with a coordinate-independent $b$. One further assume that the $r_0$ can be expanded as a series of $M/b\ll 1$
\begin{align}
r_0=b\left[c_0+c_1\dfrac{M}{b}+c_2\left(\dfrac{M}{b}\right)^2+c_3\left(\dfrac{M}{b}\right)^3+\mathcal{O}\left(\dfrac{M}{b}\right)^4\right],\label{bexpr0}
\end{align}
where $c_i$ is constant. Putting (\ref{bexpr0}) into (\ref{r0expb}), the factors $c_i$ are obtained by solving the coefficient of each term of $(M/b)^i$ is equal to zero, which are given by
\begin{align}
c_0=\sqrt{a_0},~~c_1=\dfrac{a_1}{2a_0},~~c_2=\dfrac{4a_0a_2-3a_1^2}{8\sqrt{a_0^5}},~~c_3=\dfrac{a_1^3-2a_0a_1a_2+a_0^2a_3}{2a_0^4}.
\end{align}
Putting (\ref{Arexpand}) and (\ref{Brexpand}) into (\ref{newdeflectionangle}), and then expanding the integrated function as a series in $M/r_0$ to the same order. The light deflection angle is obtain by carrying out the integration in (\ref{newdeflectionangle}) term by term
\begin{align}
\hat{\alpha}(r_0)=(\sqrt{b_0}-1)\pi+\dfrac{a_0b_1-a_1b_0}{a_0\sqrt{b_0}}\dfrac{M}{r_0}+\lambda\left(\dfrac{M}{r_0}\right)^2+\xi\left(\dfrac{M}{r_0}\right)^3+\mathcal{O}\left(\dfrac{M}{r_0}\right)^4,\label{alphar0}
\end{align}
where
\begin{align}
\lambda=&\dfrac{8a_1^2b_0^2(\pi-1)+4a_0a_1b_0b_1(\pi-2)+a_0(4a_0b_0b_2-8a_2b_0^2-a_0b_1^2)\pi}{16a_0^2\sqrt{b_0^3}},\notag\\
\xi=&\frac{1}{48 a_0^3 \sqrt{b_0^5}}\left[4 a_0^2 (a_0 [8 b_0^2 b_3-4 b_0 b_1 b_2+b_1^3]-6 a_2 b_0^2 b_1-24 a_3 b_0^3)+3a_0a_1b_0(8a_2b_0^2(9-\pi)+a_0(b_1^2-4b_0b_2)[4-\pi])\right.\notag\\
&\left.+6a_0a_1^2b_0^2b_1(9-2\pi)-2a_1^3b_0^3(67-12\pi)\right].\notag
\end{align}
Utilizing the (\ref{bexpr0}), we obtain the deflection angle of light ray in (\ref{alphar0}) as a function of $b$ as follows \footnote{Because the higher-order corrections to light deflection angle would be more complicated formulas, here we solely approximate up to the third order.}:
\begin{align}
\hat{\alpha}(b)=A_0+A_1\dfrac{M}{b}+A_2\left(\dfrac{M}{b}\right)^2+A_3\left(\dfrac{M}{b}\right)^3+\mathcal{O}\left(\dfrac{M}{b}\right)^4,\label{alphab}
\end{align}
where
\begin{align}
A_0=&(\sqrt{b_0}-1)\pi,~~A_1=\dfrac{a_0b_1-a_1b_0}{\sqrt{a_0^3b_0}},\label{oneA1}\\
A_2=&\dfrac{\pi}{16a_0^3\sqrt{b_0^3}}\left[8a_1^2b_0^2-4a_0a_1b_0b_1+a_0(4a_0b_0b_2-8a_2b_0^2-a_0b_1^2)\right],\label{twoA2}\\
A_3=&\dfrac{1}{12\sqrt{a_0^9b_0^5}}\left[15a_0a_1^2b_0^2b_1-35a_1^3b_0^3+3a_0a_1b_0[20a_2b_0^2+a_0(b_1^2-4b_0b_2)]+a_0^2(a_0[b_1^3-4b_0b_1b_2+8b_0^2b_3]\right.\notag\\
&\left.-24a_3b_0^3-12a_2b_0^2b_1)\right].\label{threeA3}
\end{align}
It is clearly seen that the first term $A_0$ on the right hand side of the (\ref{alphab}) doesn't vanish for a static, spherically
symmetric and not asymptotically flat spacetime. The generalized formalism of the deflection angle in (\ref{alphab}) will reduce to the literature's result \cite{Keeton:2005jd} in the static, spherically symmetry asymptotically Minkowski spacetime, when the $a_0=b_0=1$. Meanwhile, in terms of the (\ref{SchwarPPN}), we have examined that the our result of the deflection angle (\ref{alphab}) is consistent to the literature's result \cite{Keeton:2005jd} in the Schwarzschild black hole spacetime. In the next section, we will apply the (\ref{alphab}) to the calculation
of the weak deflection angle of light for a static, spherically symmetry not  asymptotically flat spacetime, namely the metric-affine bumblebee gravity.

\section{Gravitational lensing by a Schwarzschild-like black hole in metric-affine bumblebee gravity}
\label{section3}
In this section, we first briefly review the Schwarzschild-like black hole solution
in metric-affine bumblebee gravity, and then we calculate the weak deflection angle for a pair of asymptotically region light source and observer. In addition, we also derive the angular radius of the Einstein's ring in this theory framework, and estimate the Lorentz-violating coefficient $\ell$ from astrophysics data of the Einstein's ring.

\subsection{Schwarzschild-like metric in metric-affine bumblebee gravity}
The action of metric-affine bumblebee gravity is written as \cite{Filho:2022yrk}:
\begin{align}
S=\int d^4x\sqrt{-g}\left[\dfrac{1}{16\pi}\left[(1-u)R(\Gamma)+s^{\mu\nu}R_{\mu\nu}(\Gamma)\right]-\dfrac{1}{4}B^{\mu\nu}B_{\mu\nu}-V(B^{\mu}B_{\mu}\pm \bar{b}^2)\right]+\int d^4x\sqrt{-g}\mathcal{L}_{mat}(g_{\mu\nu},\psi),\label{bumblebeeaction}
\end{align}
where $R(\Gamma)$ and $R_{\mu\nu}(\Gamma)$ are the Ricci scalar and Ricci tensor with respect to the connection $\Gamma$, respectively; $u=\zeta B/4$ and $s^{\mu\nu}=\zeta B^{\mu}B^{\nu}$ denote coefficients (fields) responsible for the explicit (local) LSB; $B\equiv g^{\mu\nu}B_{\mu}B_{\nu}$ with the bumblebee field $B_{\mu}$, and its field strength $B_{\mu\nu}=(dB)_{\mu\nu}$. It is worth mentioning that the bumblebee field requires a suitable potential  $V(B^{\mu}B_{\mu}\pm \bar{b}^2)$ with a nonzero vacuum expectation value, inducing a spontaneous LSB, i.e. $<B_{\mu}>=\bar{b}_{\mu}$, among $\bar{b}_{\mu}$ is a minimum of the potential, and $\bar{b}^2$ is defined by $g^{\mu\nu}\bar{b}_{\mu}\bar{b}_{\nu}$;
$\mathcal{L}_{mat}(g_{\mu\nu},\psi)$ is the matter source Lagrangian with matter fields $\psi$. By variation of this action (\ref{bumblebeeaction}) about the connection $\Gamma$, the field equation is obtained as follows
\begin{align}
&\left(1-\dfrac{\zeta B^2}{4}\right)R_{(\mu\nu)}(\Gamma)-\dfrac{1}{2}g_{\mu\nu}R(\Gamma)+2\zeta [B^{\alpha}B_{(\mu}R_{\nu )\alpha}(\Gamma)]\notag\\
&-\dfrac{\zeta}{4}B_{\mu}B_{\nu}R(\Gamma)-\dfrac{\zeta}{2}g_{\mu\nu}B^{\alpha}B^{\beta}R_{\alpha\beta}(\Gamma)+\dfrac{\zeta}{8}B^2g_{\mu\nu}R(\Gamma)=8\pi T_{\mu\nu},\label{bumbfieldquation}
\end{align}
where the energy momentum tensor
\begin{align}
T_{\mu\nu}= T_{\mu\nu}^{mat}+T_{\mu\nu}^{B}
\end{align}
with
\begin{align}
T_{\mu\nu}^{mat}=&-\dfrac{2}{\sqrt{-g}}\dfrac{\rm{\delta}(\sqrt{-g}\mathcal{L}_{mat})}{\rm{\delta}g^{\mu\nu}},\label{mattertensor}\\
T_{\mu\nu}^{B}=&B_{\mu\sigma}B_{\nu}~^{\sigma}-\dfrac{1}{4}g_{\mu\nu}B^{\alpha}~_{\sigma}B^{\sigma}~_{\alpha}-Vg_{\mu\nu}+2V'B_{\mu}B_{\nu}.\label{bumbtensor}
\end{align}

For static spherically symmetric solutions to (\ref{bumbfieldquation}), one can write down the following ansatz:
\begin{align}
ds^2=-e^{2\sigma(r)}dt^2+e^{-2\rho(r)}dr^2+r^2(d\theta^2+\sin^2\theta d\phi^2),\label{ansatzmetric}
\end{align}
where $\sigma(r)$ and $\rho(r)$ are the metric functions.
According to \cite{Filho:2022yrk}, one chooses a fixed bumblebee field to assume its vacuum expectation value ($<B_{\mu}>=b_{\mu}$) compels $V=0$ and $V'=0$, and taking into account $T^{mat}_{\mu\nu}=0$. Meanwhile, using the ansatz (\ref{ansatzmetric}) together with (\ref{bumbtensor}), one then solve the field equation, which gives
\begin{align}
e^{2\sigma(r)}=\dfrac{1-\dfrac{2M}{r}}{\sqrt{\left(1+\dfrac{3\ell}{4}\right)\left(1-\dfrac{\ell}{4}\right)}},~~e^{-2\rho(r)}=\left(1-\dfrac{2M}{r}\right)^{-1}\sqrt{\dfrac{1+\dfrac{3\ell}{4}}{\left(1-\dfrac{\ell}{4}\right)^3}},\label{bumblebeemetric}
\end{align}
where the Lorentz-violating coefficient $\ell=\zeta \bar{b}^2$, and $A(r)=e^{2\sigma(r)}$ and $B(r)=e^{-2\rho(r)}$ in this paper. It is worth mentioning that the spacetime is not asymptotically
Minkowski in this case due to the nonvanishing of the Riemann tensor in the limit $r\rightarrow \infty$. For details of the above derivation, see Ref. \cite{Filho:2022yrk}.

\subsection{The light deflection angle around the Schwarzschild-like black hole}
We expand $A(r)$ and $B(r)$ in (\ref{bumblebeemetric}) as a series of the small parameter $M/r$, and then the corresponding PPN parameters are obtained
\begin{align}
a_0=&\frac{4}{\sqrt{16+8 \ell-3 \ell^2}},~a_1=-\frac{8}{\sqrt{16+8 \ell-3 \ell^2}},~a_2=a_3=0;\label{bumbPPNai}\\
b_0=&4 \sqrt{\frac{4+3 \ell}{(4-\ell)^3}},~b_1=8 \sqrt{\frac{4+3 \ell}{(4-\ell)^3}},~b_2=16 \sqrt{\frac{4+3 \ell}{(4-\ell)^3}},~b_3=32 \sqrt{\frac{4+3 \ell}{(4-\ell)^3}}.\label{bumbPPNbi}
\end{align}
Note that here $a_0\neq b_0\neq 1$ due to the presence of the $\ell$, which will lead to the zero-order term $A_0$ of the light deflection angle in (\ref{alphab}) not vanishing. Substituting these PPN parameters in (\ref{bumbPPNai}) and (\ref{bumbPPNbi}) into (\ref{oneA1})-(\ref{threeA3}), the light deflection angle in the metric-affine bumblebee gravity is given by
\begin{align}
\hat{\alpha}(b)=\left(2 \sqrt[4]{\frac{4+3 \ell}{(4-\ell)^3}}-1\right)\pi+4 \sqrt{\frac{16}{4-\ell}-3}\dfrac{M}{b}+\frac{15\pi}{8}\sqrt[4]{\frac{(4+3 \ell)^3}{4-\ell}}\left(\dfrac{M}{b}\right)^2+\frac{32(3 \ell+4)}{3}\left(\dfrac{M}{b}\right)^3+\mathcal{O}\left(\dfrac{M}{b}\right)^4.\label{bumbleangle}
\end{align}
By setting $\ell=0$, our result (\ref{bumbleangle}) reduces to the result established by GR for the bending of light \cite{Keeton:2005jd}. To clearly see the contribution of $\ell$, assuming $\ell\ll 1$, and then the deflection angle (\ref{bumbleangle}) is rewritten as \footnote{We only show approximations up to second-order for the light deflection angle to simplify following the analytic calculation of the angular radius of the Einstein’s ring to second-order approximations.}
\begin{align}
\hat{\alpha}(b)=4\dfrac{M}{b}+\frac{3 \pi  \ell}{8}+2\dfrac{\ell M}{b}+\frac{15 \pi }{4}\left(\dfrac{M}{b}\right)^2+\frac{3 \pi  \ell^2}{128}+(\text{terms of order} \geq 3).\label{bumbleanglenew}
\end{align}
In order to facilitate following sector our discussion, we assume that $\mathcal{O}(M/b) \sim \mathcal{O}(\ell) \sim \mathcal{O}(\varepsilon)\ll 1$, where $\varepsilon$ is a dimensionless small
quantity.

\subsection{The size of the Einstein ring around the Schwarzschild-like black hole}
\label{section4}
Next, we further calculate the analytical expression of the
angular radius of the Einstein’s ring in metric-affine bumblebee gravity. We merely consider the extraordinary situation that the source, lens and observer are aligned along the same axis, and the source and observer are located at asymptotically region. In addition, we estimate the upper bound of the Lorentz-violating coefficient $\ell$ by using observational data of the Eistein's ring of the galaxy ESO325-G004.

\begin{figure}[h]
\centering
\includegraphics[width=3.2in]{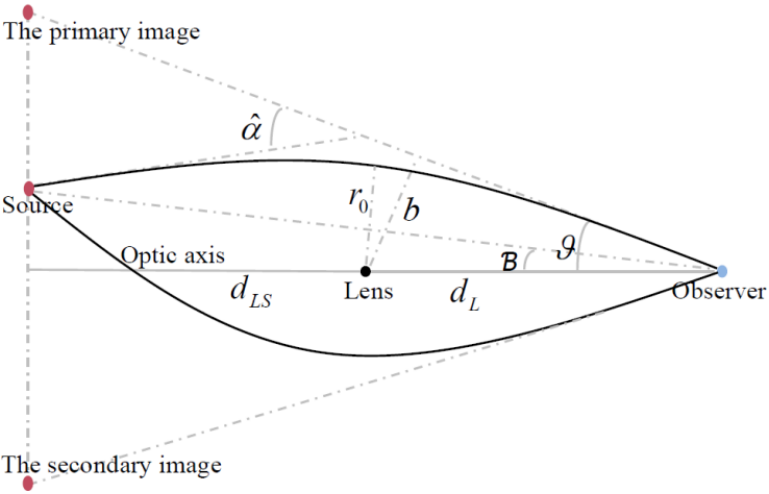}
\caption{The schematic diagram of light bending and GL \cite{Gao:2019pir}.}
\label{lenspicture}
\end{figure}

Without assuming the intersection point between the incoming and outgoing ray trajectories lies on the lens plane, Bozza derived an improved version of the lens equation as~\cite{Bozza:2008ev}
\begin{align}
d_S\tan\mathcal{B}=\dfrac{d_L\sin\vartheta-d_{LS}\sin(\hat{\alpha}-\vartheta)}{\cos(\hat{\alpha}-\vartheta)},\label{Bozzalensequation}
\end{align}
From the figure \ref{lenspicture}, $\mathcal{B}$ is the angular position of the unlensed source; $\vartheta$ is the angular position of an image;  $d_L$ is the angular diameter distance from the observer to the lens plane; $d_{LS}$ is the angular diameter distance from the lens plane to the source plane; $d_S=d_{LS}+d_L$ and $\sin\vartheta=b/d_L$.

The source, lens and observer are aligned that leads to $\mathcal{B}=0$, and thus in the weak deflection approximation\footnote{It is clearly seen that $\hat \alpha$ and $\vartheta$ are of the order $\mathcal{O}(\varepsilon)$ as $\mathcal{B}=0$ from (\ref{bumbleanglenew}) and (\ref{Bozzalensequation}).} the (\ref{Bozzalensequation}) is written as
\begin{align}
\vartheta_E= \frac{d_{LS}}{d_S}\hat{\alpha}+\mathcal{O}\left(\varepsilon^3\right),\label{Einsteinring}
\end{align}
where $\vartheta_E$ denotes the angular radius of the Einstein's ring.
Putting (\ref{bumbleanglenew}) and
\begin{align}
\vartheta_E = \arcsin\left(\dfrac{b}{d_L}\right) = \dfrac{b}{d_L} + \mathcal{O}\left(\varepsilon^3\right)
\end{align}
into (\ref{Einsteinring}), we obtain
\begin{align}
\vartheta_E =\dfrac{d_{LS}}{d_S}\left[\dfrac{4M}{d_L\vartheta_E}+\dfrac{3\pi \ell}{8}+\dfrac{15\pi}{4}\dfrac{M^2}{d_L^2\vartheta_E^2}+\dfrac{3\pi \ell^2}{128}+2\ell\dfrac{M}{d_L\vartheta_E}\right]+ \mathcal{O}\left(\varepsilon^3\right).\label{new-vartheta2}
\end{align}

In the following we obtain the $\vartheta_E$ by solving  (\ref{new-vartheta2}). From the first-order terms in (\ref{new-vartheta2}), we can derive
\begin{align}
 \vartheta_E^2=\dfrac{d_{LS}}{d_S}\dfrac{4M}{d_L}+\dfrac{d_{LS}}{d_S}\dfrac{3\pi \ell}{8}\vartheta_E +\mathcal{O}\left(\varepsilon^3\right),\label{b2DLexpression}
\end{align}
which indicates $\mathcal{O}(M/d_L)\sim \mathcal{O}(\varepsilon^2)$.
Therefore, the (\ref{new-vartheta2}) is rewritten as:
\begin{align}
\vartheta_E^3=\dfrac{d_{LS}}{d_S}(\sigma_1+\gamma_2)\vartheta_E^2+\dfrac{d_{LS}}{d_S}(\delta_2+\eta_3)\vartheta_E+\dfrac{d_{LS}}{d_S{}}\xi_4+ \mathcal{O}\left(\varepsilon^5\right),\label{simplifyvartheta}
\end{align}
where
\begin{align}
\sigma_1=\dfrac{3\pi \ell}{8},~~\gamma_2=\dfrac{3\pi \ell^2}{128},~~\delta_2=\dfrac{4M}{d_L},~~\eta_3=\dfrac{2M\ell}{d_L},~~\xi_4=\dfrac{15\pi M^2}{4d_L^2},\label{threecoefficients}
\end{align}
among each number in the subscript represents the order, i.e.\ $\sigma_1\sim\mathcal{O}(\varepsilon), \gamma_2\sim \delta_2\sim\mathcal{O}(\varepsilon^2)$, $\eta_3\sim\mathcal{O}(\varepsilon^3)$ and $\xi_4\sim\mathcal{O}(\varepsilon^4)$.
We assume that $\vartheta_E$ is written in different orders as:
\begin{align}
\vartheta_E=\vartheta_{E1}+\vartheta_{E2}+\mathcal{O}\left(\varepsilon^3\right),\label{expandvartheta}
\end{align}
then we can solve (\ref{simplifyvartheta}) order by order, i.e.\ solve
\begin{align}
\vartheta_{E1}^3=&\dfrac{d_{LS}}{d_S}(\delta_2+\sigma_1\vartheta_{E1})\vartheta_{E1},\label{equation1}\\
3\vartheta_{E1}^2\vartheta_{E2}=&\dfrac{d_{LS}}{d_S}[\delta_2\vartheta_{E2}+\xi_4+\vartheta_{E1}(\eta_3+\gamma_2\vartheta_{E1}+2\sigma_1\vartheta_{E2})],\label{eqyation2}
\end{align}
which gives
\begin{align}
\vartheta_{E1}=&\dfrac{d_{LS}\sigma_1+\sqrt{4d_Sd_{LS}\sigma_2+d_{LS}^2\sigma_1^2}}{2d_S},\label{E1value}\\
\vartheta_{E2}=&\dfrac{1}{2}\left[\dfrac{\xi_4}{\delta_2}+\dfrac{\sqrt{d_{LS}}\left(2d_S\delta_2\eta_3+d_{LS}\gamma_2\delta_2\sigma_1-d_S\xi_4\sigma_1+\gamma_2\delta_2\sqrt{4d_Sd_{LS}\delta_2+d_{LS}^2\sigma_1^2}\right)}{d_S\delta_2\sqrt{4d_S\delta_2+d_{LS}\sigma_1^2}}\right].\label{E2value}
\end{align}
Finally, combining (\ref{E1value}) and (\ref{E2value}) with (\ref{threecoefficients}), the analytical expression of the angular radius of the Einstein ring (\ref{expandvartheta}) is obtained
\begin{align}
\vartheta_E=&\sqrt{\dfrac{4d_{LS}M}{d_Sd_L}+\dfrac{9\pi^2d_{LS}^2\ell^2}{256d_S^2}}+\dfrac{3\pi d_{LS} \ell}{16d_S}+\dfrac{1}{256}\notag\\
&\times\left[\dfrac{120\pi M}{d_L}+\dfrac{3d_{LS}\pi \ell^2}{d_S}+\dfrac{8\sqrt{d_{LS}}(512-45\pi^2)M\ell}{d_L\sqrt{\dfrac{1024d_SM}{d_L}+9d_{LS}\pi^2\ell^2}}+\dfrac{9\sqrt{d_{LS}^3}\pi^2\ell^3}{d_S\sqrt{\dfrac{1024d_SM}{d_L}+9d_{LS}\pi^2\ell^2}}\right]+ \mathcal{O}\left(\varepsilon^3\right),\label{vartheta-value}
\end{align}
\subsubsection{Estimation of the LSB coefficient $\ell$ from the
observational data of the galaxy ESO325-G004}
Taking the galaxy ESO325-G004 as the lens, our aim here is to estimate upper bound of the $\ell$ from the observational data of Einstein' ring. All the data below are from \cite{Smith:2005pq,Smith:2013ena}, and we use $G\neq 1$ and $c\neq 1$ in this sector  analysis and calculation.

\begin{figure}[h]
\centering
\includegraphics[width=3.9in]{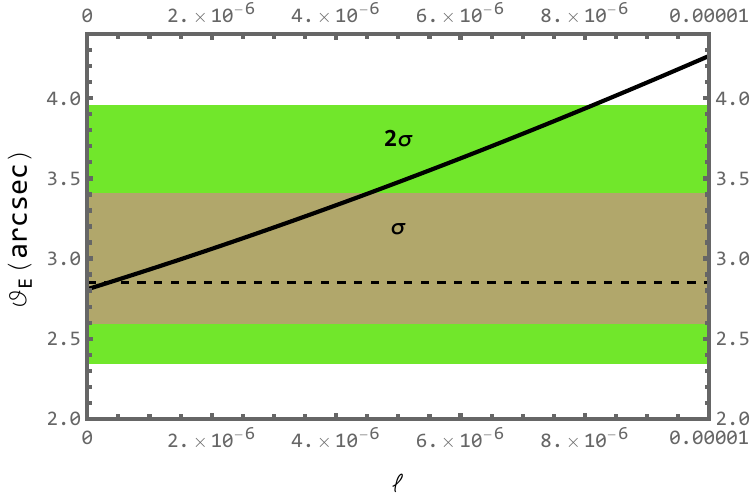}
\caption{The Einstein ring as a function of the LSB coefficient $\ell$. The angular radius of the Einstein's ring estimated with the Hubble Space Telescope observations, and its uncertainties in $1\sigma$ (or $2\sigma$) confidence levels are marked with a horizontal dashed-line and dark gray (or green) shaded regions.}
\label{EinsteinX}
\end{figure}
The observed value of the Einstein ring by lensing by the elliptical galaxy ESO325-G004 at $z_l=0.0345$ is
\begin{align}
\vartheta_E^{obs}=(2.85_{-0.25}^{+0.55})''.\label{obsering}
\end{align}
The mass of the galaxy
ESO325-G004 is $M=1.50\times 10^{11} M_{\odot}$, where $M_{\odot}=1.98\times 10^{30}{\rm kg}$ is the mass of the Sun\footnote{Here the total mass $M$ inside the Einstein's ring of the galaxy ESO325-G004 not only includes black hole mass, but also incorporates the mass of dark matter and luminous matter.}. The redshift of the background source galaxy $z_s=1.141$. Hubble has found that the relation between the spectrum redshift and distance of galaxy i.e. $cz=H_0 D$, where $H_0=70.4\, {\rm km}\, {\rm s}^{-1}\,({\rm Mpc})^{-1}$ is the Hubble constant and $D$ denotes the proper distance, related to the comoving distance $d$ by $d=D(1+z)$ for a flat cosmology. Then we have
\begin{align}
d_S=\dfrac{c z_s(1+z_s)}{H_0}=2.863\times10^4\,{\rm Mpc}, \quad d_L=\dfrac{c z_l(1+z_l)}{H_0}=1.542\times10^2\,{\rm Mpc} \ .\label{ESOdata}
\end{align}

According to the above data, the angular radius of the Einstein ring (\ref{vartheta-value}) as a function of the $\ell$ is shown in figure \ref{EinsteinX}. It is found that the $\vartheta_E$ increases gradually with the LSB coefficient $\ell$ increases. From the figure \ref{EinsteinX}, we clearly see that $\ell$ can be constrained as $\ell\lesssim 4.472\times 10^{-6}$ within $1\sigma$ uncertainties and $\ell\lesssim 8.084\times 10^{-6}$ within $2\sigma$ uncertainties. In addition, by using the average error formula $(\sum|\ell_i-\bar{\ell}|/n)$, we do every rough with error analysis, where $\ell_i$ denotes every measured value, $\bar{\ell}$ represents the average value, and $n$ is the number of measured value. Finally, the average error with $\ell\simeq4.472\times 10^{-6}$ and $8.084\times 10^{-6}$ is about $1.806\times 10^{-6}$.

\section{Shadow by a Schwarzschild-like black hole
in metric-affine bumblebee gravity}
\label{section4}
In this section, we investigate the influence of the LSB coefficient $\ell$ on the black hole shadow in metric-affine bumblebee gravity. By taking the ${\rm M}87$ galaxy as the lens, we can expect to set an upper limit on $\ell$ with the shadow of the supermassive black hole in the ${\rm M}87$ galaxy detected by the EHT.

\begin{figure}[h]
\centering
\includegraphics[width=3.2in]{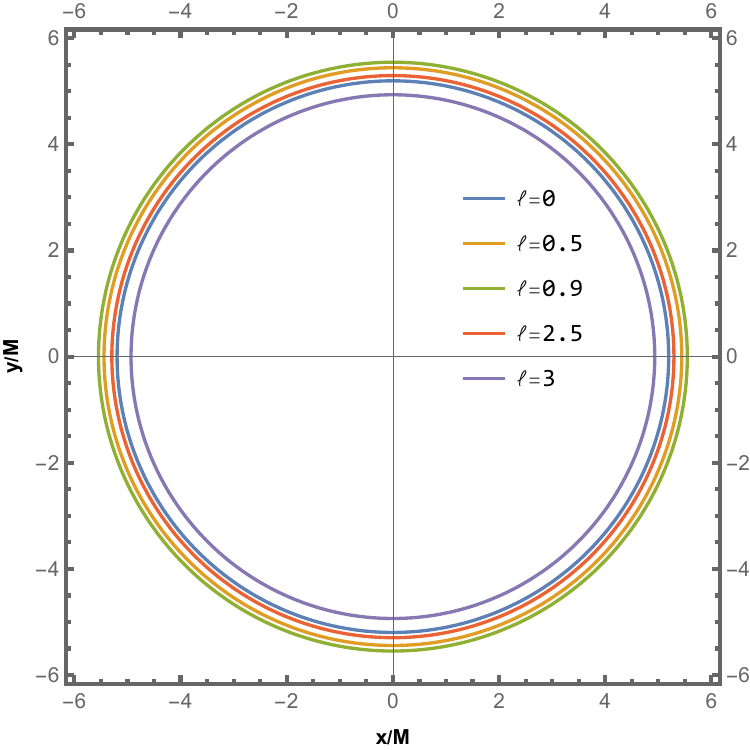}
\caption{Boundary of the black hole shadow for different values of the LSB coefficient $\ell$.}
\label{bumbleshadow}
\end{figure}
For an observer at asymptotically region, the shadow shape is described by the celestial coordinates $x$ and $y$ as follows \cite{Bardeen1973,Kuang:2022xjp}:
\begin{align}
x=\lim_{r_o \to \infty}\left(-r_o^2\sin\theta_o\dfrac{d\phi}{dr}\right),~~y=\lim_{r_o \to \infty}\left(r_o^2\dfrac{d\theta}{dr}\right),\label{celcoor}
\end{align}
in which $r_o$ is the observer distance to the black hole, $\theta_o$ is the inclination angular of the observer. We restrict our study to the equatorial plane
$\theta_o=\pi/2$, hence the radius of the shadow is given by
\begin{align}
R_s=\sqrt{x^2+y^2}=b_c,
\end{align}
where $b_c$ is the critical impact parameter corresponded to the radius of the black hole shadow. From (\ref{impactb}), the $b_c$ is obtained in metric-affine bumblebee gravity
\begin{align}
b_c=\frac{3}{2} \sqrt{3} M \sqrt[4]{-3 \ell^2+8 \ell+16},\label{bumbleB}
\end{align}
which reduces to the Schwarzschild black hole case as $\ell\to 0$. In figure \ref{bumbleshadow}, we show the shadow boundaries of the black hole for different values of $\ell$. It obviously shows that the shape of the black hole shadow is a perfect circle, and the shadow size first increases and then decreases when $\ell$ increases.

\begin{figure}[h]
\centering
\includegraphics[width=3.6in]{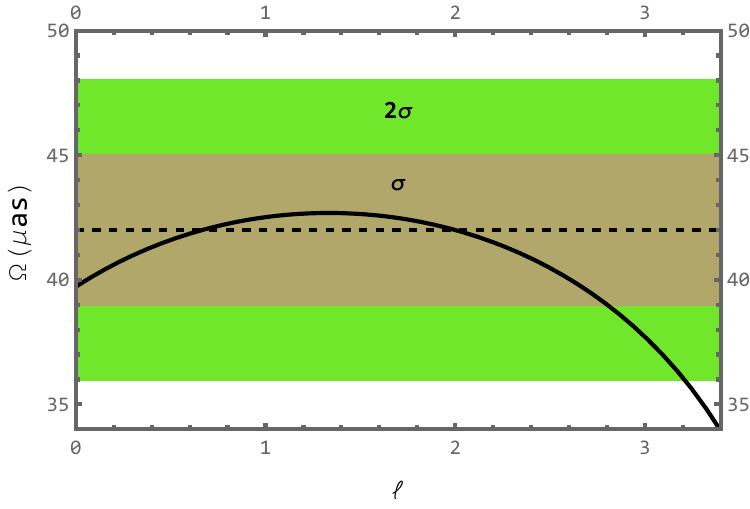}
\caption{The black curve plots the angular size of the black hole shadow as
a function of the LSB coefficient $\ell$. The
angular size of the black hole shadow for the ${\rm M}87$ galaxy estimated with the
EHT observations, and its uncertainties in $1\sigma$ (or $2\sigma$) confidence
levels are marked with a horizontal dashed-line and dark
gray ( or green) shaded regions.  }
\label{bumbledshX}
\end{figure}
It is worth mentioning that the measurements of the shadow size around the black hole may help to estimate the black hole parameters. The literatures \cite{EventHorizonTelescope:2019dse,EventHorizonTelescope:2019ths} reported that the angular size of the black hole shadow in the ${\rm M}87$ galaxy detected by the EHT is $\Omega=(42\pm 3) \mu as$, its distance is $r_o=16.8^{+0.8}_{-0.7}
{\rm Mpc}$, and its black hole mass $M=(6.5\pm 0.9)\times 10^{9} M_{\odot}$. Assuming $r_o\gg 2M$ and $\ell$, the relation between the angular size $\Omega$ and the diameter $d_{sh}$ for black hole shadow is written as
\cite{Bambi:2019tjh,Allahyari:2019jqz}:
\begin{align}
\Omega\simeq \dfrac{d_{sh}}{r_o},\label{dshandangular}
\end{align}
where $d_{sh}\simeq2b_c$. From the (\ref{dshandangular}), we have plotted the angular size of the black hole shadow for the metric-affine bumblebee gravity as a function of the $\ell$ in figure \ref{bumbledshX}. As shown in figure \ref{bumbledshX}, $\ell$ is upper bounded by the constraint $\ell\lesssim 2.801$ at $1\sigma$ and $\ell\lesssim 3.207$ at $2\sigma$. Similar to the previous example, by utilizing $(\sum|\ell_i-\bar{\ell}|/n)$, the average error is about $0.203$ for $\ell\simeq2.801$ and $2.307$.

\section{Summary and Conclusion}
\label{section5}
In this paper, we have derived a generalized PPN formalism of the light weak deflection angle in a static, spherically symmetric and not asymptotically flat spacetime, and then applied this general expression to metric-affine bumblebee gravity model. Moreover, we analytically calculated the light weak deflection angle and the angular radius of Einstein's ring of this background within the weak-field approximation, and analyzed the black hole shadow in this theory framework. Meanwhile, using the astrophysical observational data, we roughly constrained the upper bound of the LSB coefficient $\ell$.

The metric-affine bumblebee gravity model, which introduces a bumblebee vector field that couples to spacetime curvature, provide us a unique perspective on gravitational interactions. The bumblebee vector field requires the presence of the potential  $V(B^{\mu}B_{\mu}\pm \bar{b}^2)$ with a nonzero vacuum expectation value induces a spontaneous LSB. Since LSB can provide candidate low energy signals at the Planck scale for a unified quantum theory of gravity and other forces, the influences of LSB on various astrophysical phenomena should be extensively investigated and attended in cosmological and astrophysical communities. In present work, we have provided our findings that demonstrated the significant influence of the LSB on several observables, such as the light deflection angle, the Einstein's ring and the black hole shadow. We have found that the light deflection angle and the Einstein's ring gradually increase with increasing of
the LSB coefficient $\ell$, while the shadow size first increases and then decreases as $\ell$ increases. Meanwhile, we obtained the upper bounds for the LSB coefficient $\ell\lesssim 4.472\times 10^{-6}$ within $1\sigma$ uncertainties and $\ell\lesssim 8.084\times 10^{-6}$ within $2\sigma$ uncertainties, from the astrophysical data of the Einstein's ring. Furthermore, utilizing the most recent observational data from the shadow of the supermassive black hole in the ${\rm M}87$ galaxy, the $\ell$ relied on the stringent constraint $\ell\lesssim 2.081$ at $1\sigma$ and $\ell\lesssim 3.207$ at $2\sigma$. The two upper bounds of the $\ell$ from the observational data of the galaxy ESO325-G004 are obtained together with considering the total mass of black hole, dark matter and luminous matter, while the two upper bounds of the $\ell$ from the observational data of the M87 galaxy are given by merely taking into account black hole mass. Moreover, by the average error analysis, the average error of the $\ell$ for the Einstein's ring scenario is much less than that of the black hole shadow angular size as well. Therefore, the constraint values of the $\ell$ for the galaxy ESO325-G004 scenario are more significant than those for the M87 galaxy case. These findings presented in this work may offer to valuable insights into the intricate interplay between metric-affine bumblebee gravity and astrophysical phenomena, shedding light on the potential detectability of
bumblebee parameters and their impact on observables.

\section{Acknowledgements}
Thank you very much the anonymous referee for the careful review of the manuscript and providing us important advices for improvement.
I am grateful to Profs. Ya-Peng Hu and Jun-Jin Peng for their helpful discussions as well. 

\end{document}